\documentclass[11pt,twoside]{article}
\usepackage{asp2010}

\resetcounters

\begin{document}

\title{Peculiar isolated neutron stars and the source in the Carina Nebula}
\author{Adriana~Mancini~Pires
\affil{Leibniz-Institut f\"ur Astrophysik Potsdam (AIP), Germany}
}

\begin{abstract}
The new results of our observing campaign targeting the isolated neutron star 2XMM~J104608.7-594306 in the Carina Nebula are used to understand how peculiar groups of isolated neutron stars relate to each other, as well as to the bulk of the normal radio pulsar population.
\end{abstract}

\altsubsubsection*{Populations of peculiar isolated neutron stars}

The observed population of neutron stars is dominated by radio pulsars. In recent years, however, different observational manifestations of isolated neutron stars (INSs) have been discovered, which include magnetars, X-ray dim isolated neutron stars (XDINS), rotating radio transients (RRATs), and central compact objects in supernova remnants (CCOs).
While fewer in number, the peculiar INS subgroups might represent a considerable fraction of the neutron stars in the Galaxy and are very important for a number of reasons. First, the investigation of individual sources can greatly impact our understanding of the physics of matter at extreme conditions of gravity and magnetic field. Second, the relations between the various subgroups help us to understand the main properties determining the neutron star phenomenology. Finally, as new objects are discovered, a comprehensive picture of neutron star birth and evolution in the Milky Way can be aimed for.

\altsubsubsection*{The source 2XMM~J104608.7-594306}

We recently reported the results of an observational campaign with XMM-Newton and ESO-VLT to study the properties of the radio-quiet, thermally emitting INS 2XMM~J104608.7-594306 (J1046, for short), likely located in the Carina Nebula \citep{pir12}.
The observations reveal a unique type of object. The source has a spectrum that is thermal and soft, with no evidence of magnetospheric emission. Significant deviations (absorption features) from a simple blackbody model are identified. Furthermore, the optical counterpart is fainter than $m_{\rm V}=27$ ($2\sigma$) and no $\gamma$-ray emission is significantly detected from either the Fermi or AGILE data.
Very interestingly, while these characteristics are remarkably similar to those of XDINS or the only RRAT (J1819-1458, \citealt{lau07}) so far detected in X-rays, all with spin periods of a few seconds, we found intriguing evidence of very rapid rotation, $P\sim18.6$\,ms ($p_f\sim14\%$), at the $4\sigma$ confidence level.

\altsubsubsection*{Thermally emitting isolated neutron stars and missing links}

We compare the spectral and timing properties of possible ``missing links'' -- J1046, RRAT~J1819-1458 and Calvera \citep{rut08,zan11} -- with those of other thermally emitting INSs, namely XDINS, magnetars, CCOs, and millisecond pulsars (MSPs). As evidenced in the $L_{\rm X}$\,--\,$R_{\rm bb}$ and $R_{\rm bb}$\,--\,$P$ diagrams in Fig.~\ref{fig_diag}, J1046's rapid rotation is surprisingly at odds with the neutron star purely thermal energy distribution. 

\articlefigure{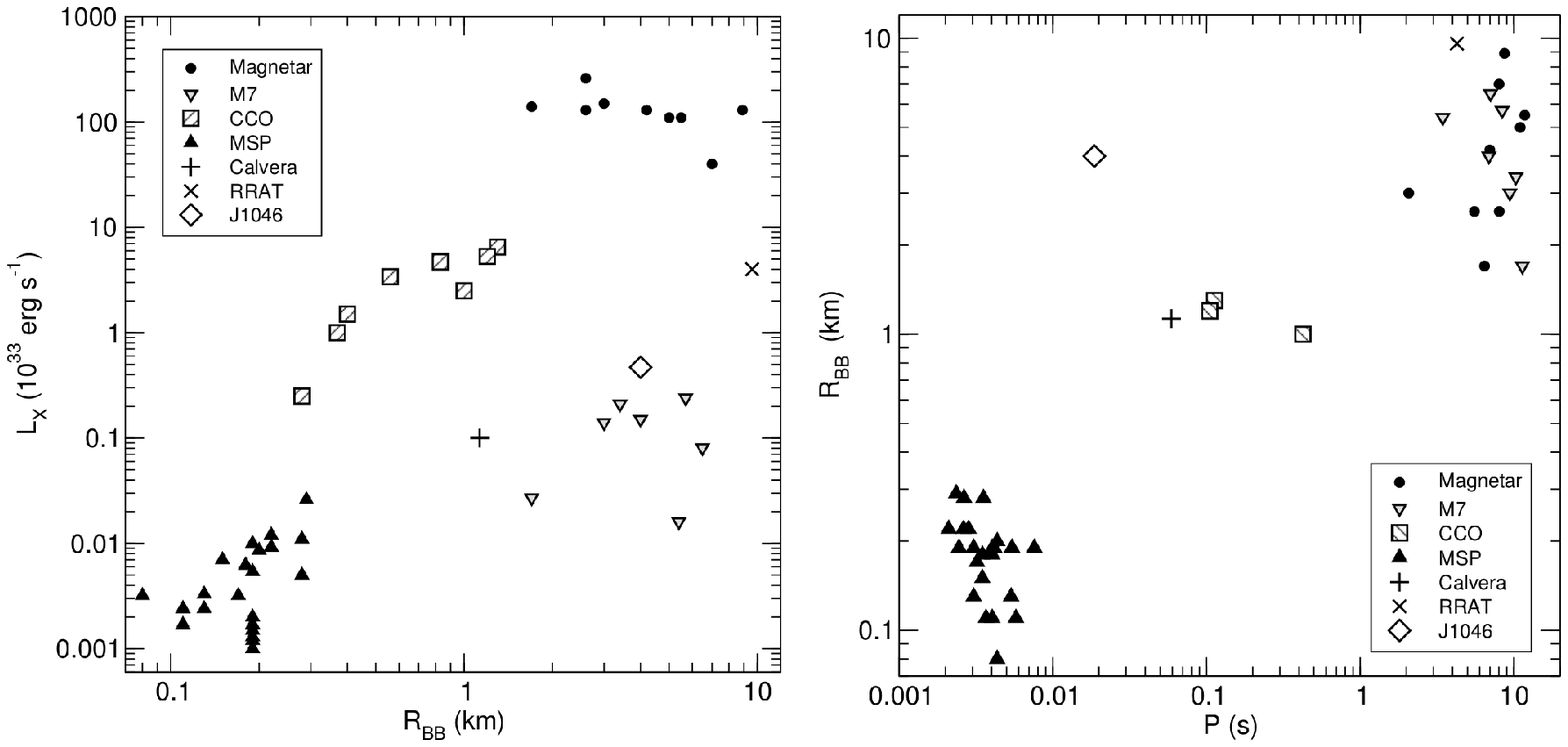}{fig_diag}{Blackbody luminosity vs. emission radius and emission radius vs. neutron star spin period for several populations of thermal INSs (see text).}

\altsubsubsection*{Possible evolutionary state}

In general, similarly fast-spinning neutron stars are expected to be either young and energetic spin-powered pulsars or, conversely, old recycled neutron stars, with low magnetic fields and large characteristic ages. The very thermal X-ray spectrum of J1046 is clearly at odds with that of a typical young spin-powered pulsar. On the other hand, the presence of the source in the Carina Nebula excludes an old, fully recycled, object. 

Keeping in mind the low statistics and given the still poorly known overall characteristics of the population of Galactic anti-magnetars \citep{hal10}, the possibility of an old CCO (as discussed for the case of Calvera) remains open for J1046, although we note that the overall spectral properties of the source do not match those expected for this scenario (e.g. double thermal component, size of emission radius, and pulsed fraction).
Furthermore, it is commonly believed that the strong thermal emission of XDINS and RRAT~J1819-1458 is a result of additional heating from magnetic field decay. The fact that J1046 shows a very similar spectrum as these sources but has not experienced significant spin down is intriguing, and shows that the magneto-rotational evolution of neutron stars is still far from being well understood.

\acknowledgements 

This work is supported by the Deutsche Forschungsgemeinschaft (grant PI~983/1-1). A.~M.~P. acknowledges work collaborators C.~Motch, R.~Turolla, A.~Schwope, M.~Pilia, A.~Treves, S.~B.~Popov and E.~Janot-Pacheco.

\bibliography{proceedings}

\begin{thebibliography}{}
\expandafter\ifx\csname natexlab\endcsname\relax\def\natexlab#1{#1}\fi
\expandafter\ifx\csname url\endcsname\relax
  \def\url#1{\texttt{#1}}\fi
\expandafter\ifx\csname urlprefix\endcsname\relax\def\urlprefix{URL }\fi
\providecommand{\eprint}[2][]{\url{#2}}

\bibitem[{{Halpern} \& {Gotthelf}(2010)}]{hal10}
{Halpern}, J.~P., \& {Gotthelf}, E.~V. 2010, ApJ, 709, 436

\bibitem[{{McLaughlin} et~al.(2007){McLaughlin}, {Rea}, {Gaensler},
  {Chatterjee}, {Camilo}, {Kramer}, {Lorimer}, {Lyne}, {Israel}, \&
  {Possenti}}]{lau07}
{McLaughlin}, M.~A., {Rea}, N., {Gaensler}, B.~M., {Chatterjee}, S., {Camilo},
  F., {Kramer}, M., {Lorimer}, D.~R., {Lyne}, A.~G., {Israel}, G.~L., \&
  {Possenti}, A. 2007, ApJ, 670, 1307

\bibitem[{{Pires} et~al.(2012){Pires}, {Motch}, {Turolla}, {Schwope}, {Pilia},
  {Treves}, {Popov}, \& {Janot-Pacheco}}]{pir12}
{Pires}, A.~M., {Motch}, C., {Turolla}, R., {Schwope}, A., {Pilia}, M.,
  {Treves}, A., {Popov}, S.~B., \& {Janot-Pacheco}, E. 2012, A\&A, in press

\bibitem[{{Rutledge} et~al.(2008){Rutledge}, {Fox}, \& {Shevchuk}}]{rut08}
{Rutledge}, R.~E., {Fox}, D.~B., \& {Shevchuk}, A.~H. 2008, ApJ, 672, 1137

\bibitem[{{Zane} et~al.(2011){Zane}, {Haberl}, {Israel}, {Pellizzoni},
  {Burgay}, {Mignani}, {Turolla}, {Possenti}, {Esposito}, {Champion},
  {Eatough}, {Barr}, \& {Kramer}}]{zan11}
{Zane}, S., {Haberl}, F., {Israel}, G.~L., {Pellizzoni}, A., {Burgay}, M.,
  {Mignani}, R.~P., {Turolla}, R., {Possenti}, A., {Esposito}, P., {Champion},
  D., {Eatough}, R.~P., {Barr}, E., \& {Kramer}, M. 2011, MNRAS, 410, 2428

\end{thebibliography}

\end{document}